\documentclass[aps,pra,twocolumn,floatfix,nofootinbib]{revtex4}

\usepackage{psfrag,graphicx}
\usepackage{dcolumn}
\usepackage{amsmath,amssymb}
\usepackage{bm}

\newcommand{\be}{\begin{equation}}
\newcommand{\ee}{\end{equation}}
\newcommand{\bq}{\begin{eqnarray}}
\newcommand{\eq}{\end{eqnarray}}

\def\be{\begin{equation}}
\def\ee{\end{equation}}
\def\bea{\begin{eqnarray}}
\def\eea{\end{eqnarray}}

\def\1/2{\frac{1}{2}}

\begin{document}

\title{Single atom quantum walk with 1D optical superlattices}

\author{Jaewoo Joo$^1$}
\author{P. L. Knight$^1$}
\author{Jiannis K. Pachos$^2$}
\affiliation{$^1$ Blackett Laboratory, Imperial College London,
Prince Consort Road, London, SW7 2BW, United Kingdom \\
$^2$Centre for Quantum Computation, Department of Applied Mathematics and
Theoretical Physics, University of Cambridge, Cambridge CB3 0WA, United
Kingdom}
\date{\today}

\begin{abstract}

A proposal for the implementation of quantum walks using cold atom
technology is presented. It consists of one atom trapped in time
varying optical superlattices. The required elements are presented
in detail including the preparation procedure, the manipulation
required for the quantum walk evolution and the final measurement.
These procedures can be, in principle, implemented with present
technology.

\end{abstract}

%\pacs{}

\maketitle

\section{Introduction}
\label{intro}

Ever since the idea of quantum walk was originally proposed in
1993 \cite{Aharonov}, it has been understood as an interference
phenomenon of quantum states \cite{Knight1}. Various
implementations of quantum walks have been proposed employing
atoms \cite{Dur}, ions \cite{Ma} or optics elements
\cite{Knight1}. One proposal \cite{Dur} which is particularly
noteworthy proposes a way to perform quantum walks in optical
lattices \cite{Jaksch04}, where cold atoms are utilized in the
Mott insulator phase. One of the main challenges phased by these
models is the decoherence of internal states in optical lattices
during quantum walks.

Here we propose an alternative scheme for implementing quantum
walks based on superlattices, which are formed by two optical
lattices possessing different frequencies \cite{Kay}. For that we
employ a single atom trapped in superpositions of time modulated
optical lattices in a one dimensional (1D) configuration. By
employing two different superlattices, the state of the atom
penetrates the left or right side through the optical potentials.
The evolution is governed by direct tunneling of the atom between
neighboring sites which can be made large enough to allow a
significant number of steps within decoherence time. The
preparation of the physical setup, the manipulation procedures and
the subsequent measurements, is presented in detail in what
follows.

In this paper, we begin with describing double well potentials in
superlattices as a building block. In order to produce step
operation in quantum walks, the state-dependent tunneling of the
atom occurs into the direction of a lower optical barrier in the
double well potential. By alternating two superlattices depending
on the number of steps, we generate a standard Hadamard driven
quantum walk. In addition, we consider the physical procedures
that are needed to trap a single atom in a 1D optical lattice. We
estimate the experimental parameters necessary for the realization
of the quantum walk and the unitary errors in step operations due
to uncontrolled laser properties. Finally, a measurement procedure
is proposed that unambiguously distinguishes the quantum evolution
from its classical counterpart.

\section{Quantum walks with superlattices}
\label{Construction}

\subsection{Superlattices as double wells}

Let us consider a single atom that encodes a qubit in two possible
states $0$ and $1$ characterized, e.g. by two different hyperfine
levels. It is possible to trap this atom at a particular site
within a 1D optical lattice that consists of two optical standing
waves. A 1D optical lattice is given by
\begin{eqnarray} \label{OL01}
V (x) &=& V \cos {2 \pi \over \lambda} x \, ,
\end{eqnarray}
where $\lambda$ is the periodicity of the optical lattice used to
trap cold atoms, viewed as dipoles, at its potential minima.

\begin{center}
\begin{figure}[b]
\resizebox{!}{7cm} {\includegraphics{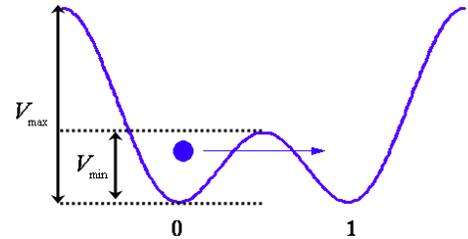}} \vspace{-4cm}
\caption{ \label{DWell} Tunneling of an atom in state
$|0\rangle_{\rm int}$ appears from the left site to the right site
in a double well ($V_{\rm max} = V + V'$ and $V_{\rm min} = V -
V'$).}
\end{figure}
\end{center}

The superlattices are generated by interference of two optical
lattices which have different frequencies. In particular, we
consider an additional lattice with potential amplitude $V'$, with
double the wavelength of the first lattice. The resulting
superlattice potential is given by
\begin{eqnarray} \label{SOL01}
V (x) &=& V \cos  {2 \pi \over \lambda} x - V' \cos
\,{\pi \over \lambda} x,~~~~
\end{eqnarray}
as shown in Figure \ref{DWell}.

Our aim is to activate tunneling from one side to the other side
depending on the internal state of the atom \cite{Kay}. As shown
in Figure \ref{DWell}, an initialized atom in the internal state
$|0 \rangle_{\rm int}$ is trapped in the left position $|0
\rangle_{x}$ where subindexes ``int'' and ``$x$'' denote an
internal state and a position state of the atom. In this section,
we assume the ideal case such that the maximum amplitude of the
superlattices $V_{\rm max}$ produces no tunneling for example
between sites labeled 0 and $-1$ (to the left of the sites shown
in Fig. \ref{DWell}) and the minimum amplitude $V_{\rm min}$
enables a tunneling between two positions in the double well.
Thus, one can restrict between two sites where the tunneling
interaction $J$ is activated by the superlattice
\begin{eqnarray} \label{SimpleV}
H = -J (a_{1}^\dagger a_{2}^{} + a_{2}^\dagger a_{1}^{}).
\end{eqnarray}
The time evolution operator is described by
\begin{eqnarray} \label{U}
U (t_1,t_0) &=& \cos \big( {\textstyle {{1 \over 2}}} J
\Delta t \big) {\bf 1} + {\rm i} \, \sin \big( {\textstyle {{1
\over 2}}} J \Delta t  \big) \sigma_x \, ,
\end{eqnarray}
where $\Delta t = t_1 - t_0$ and $\sigma_x$ is one of the Pauli
operators in the basis $|0\rangle_x$ and $ |1\rangle_x$ that
denote the two possible position states of the double well. In
order to obtain perfect tunneling between two sites, a time
$\Delta t = \pi/ J$ is required during which the initial position
state $|0\rangle_x$ becomes $|1\rangle_x$.

\subsection{Quantum walks using 1D superlattices}
\label{Quantum01}

\begin{figure}[h]
\centering
\begin{minipage}{\columnwidth}
\begin{center}
\resizebox{\columnwidth}{!}{\rotatebox{0}{\includegraphics{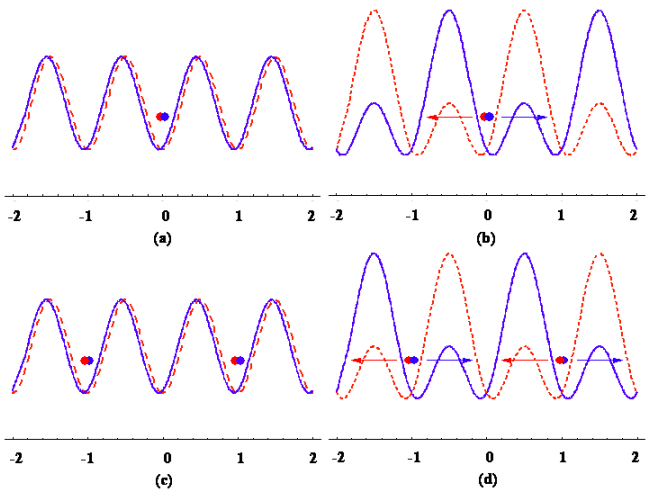}
}}
\end{center}
\vspace{0cm}  \caption{ \label{1st} (a) The superposed atom is
located at site $x=0$ of optical lattices. (b) By employing two
superlattices the tunneling is activated depending on the internal
state of the atom. (c) A Hadamard operation is performed in
optical lattices. (d) The tunneling by alternating two
superlattices occur among five sites at the second step of quantum
walks.}
\end{minipage}
\end{figure}

We next demonstrate that quantum walks can be achieved by
tunneling in two state-dependent super-lattices. In order to
describe symmetric quantum walks \cite{Dur}, we prepare an
initialized atom in superposed state $|\psi_{0} \rangle = {1\over
\sqrt{2}} (|0 \rangle_{\rm int} + {\rm i} |1 \rangle_{\rm int})
\otimes |0\rangle_x$. The subindex $n$ of state $|\psi_{n}
\rangle$ indicates the number of steps. To implement a quantum
walk we need to construct the step evolution operator \cite{Dur}
\begin{eqnarray} \label{S}
S =\sum_{l=-\infty}^{\infty} &&( |  0 \rangle_{\rm int} \langle 0|
\otimes |l+1 \rangle_x \langle l | \nonumber \\ && + | 1
\rangle_{\rm int} \langle 1| \otimes |l-1 \rangle_x \langle l |
\big).
\end{eqnarray}
As shown in Figure \ref{1st} (b) and (d), this is achieved by
alternating between two different superlattice configurations.
The first configuration of the superlattices is given by
\begin{eqnarray} \label{SOL02}
V_{\rm odd} (x) &=& V_i \cos ( 2 \pi x )  +(-1)^{i} V'_i \cos \, (
\pi x) ,~~~~
\end{eqnarray}
where $i=0,1$ corresponds to the lattice that traps the $i$-th
state of the atom. In the next step, the applied superlattices are
given by
\begin{eqnarray} \label{SOL03}
V_{\rm even} (x) &=& V_i \cos ( 2 \pi x)  -(-1)^{i} V'_i \cos \, (
\pi x) .~~~~
\end{eqnarray}

As we see in Figure \ref{1st}, the atom is located in even positions for
even $n$ while it sits in odd positions for odd $n$. According to Eq.
(\ref{U}), alternating superlattices build two different step operators
acting on the atom during the quantum walks as follows
\begin{eqnarray} \label{T_S3}
{S}_{\rm odd} &=& \sum_{l=-\infty}^{\infty}
 \big ( | 0 \rangle_{\rm int} \langle 0| \otimes |2l+1\rangle_x
 \langle 2l|  \nonumber \\
&& ~~~~~+ | 1 \rangle_{\rm int} \langle 1| \otimes  |2l-1
\rangle_x  \langle 2l | \big), ~~~~~ \\
{S}_{\rm even} &=& \sum_{l=-\infty}^{\infty}
 \big ( | 0 \rangle_{\rm int} \langle 0| \otimes
|2l \rangle_x \langle 2l-1 | \nonumber \\
&& ~~~~~+ | 1 \rangle_{\rm int} \langle 1| \otimes |2l-2\rangle_x
\langle 2l-1 | \big). ~~~~~
\end{eqnarray}
Indeed, by applying these potentials in succession, two step
operators perform one-directional state-dependent walks
identically and the step operator $S$ can be implemented.

In order to generate a superposed internal state in every step, we
introduce the Hadamard operation \cite{rev}
\begin{eqnarray} \label{Hada}
H= {1 \over \sqrt{2}} \left( \begin{array}{cc}
1 & 1 \\
1 & - 1
\end{array} \right),
\end{eqnarray}
in the basis $|0 \rangle_{\rm int}$ and $|1 \rangle_{\rm int}$.
When this operation is performed on all atoms in optical lattices,
the internal states in each site evolve into superposition states.
In terms of quantum optics, this Hadamard operator is decomposed
into three Pauli operators which can be achieved by sequential
$\pi/2$ laser pulses over all the sites (see Section IIIB in Ref.
\cite{Dur}). Combined with the Hadamard operation between every
step operator the standard Hadamard quantum walk is implemented.
After performing $n$ steps, the final state is given by $|\psi_n
\rangle = (S H)^n |\psi_0 \rangle$.

We start to perform the Hadamard operation on the initialized
state $|\psi_0 \rangle$. After the first tunneling occurs during
time $\Delta t = \pi/ J_{\rm min}$, the state of the atom is
described by
\begin{eqnarray} \label{1st01}
|\psi_{1}\rangle &=& {1 \over \sqrt{2}} ( {\rm e}^{\rm i {\pi
\over 4}} | 0 \rangle_{\rm int} \otimes | 1 \rangle_{x} + {\rm
e}^{-\rm i {\pi \over 4}}|1 \rangle_{\rm int} \otimes | - 1
\rangle_{x}).~~~~~
\end{eqnarray}
As a result, the state of the atom becomes an entangled state
between the internal states and their positions. Then, we switch
off the two additional lasers $V'_{i}$ ($i=0,1$), which
effectively turns off the tunneling between the two neighboring
sites. If the Hadamard operation is applied on three sites, the
atomic state is described in terms of both internal states and
positions (see Figure \ref{1st} (c)).

To complete the second step of the quantum walk, we perform the
step operation by the superlattice ${V}_{\rm even} (x)$ as shown
in Figure \ref{1st} (d). Then, the atomic state after the second
step of the walk equals
\begin{eqnarray} \label{2nd01}
|\psi_{2}\rangle &=& {1 \over 2} \big( | 0 \rangle_{\rm int}
\otimes  ( {\rm e}^{\rm i {\pi \over 4}} | 2 \rangle_{x} + {\rm
e}^{-\rm i {\pi \over 4}} | 0 \rangle_{x}) \nonumber \\ &&\, + |1
\rangle_{\rm int} \otimes  ( {\rm e}^{\rm i {\pi \over 4}} | 0
\rangle_{x} - {\rm e}^{-\rm i {\pi \over 4}} | -2 \rangle_{x})
\big).~~
\end{eqnarray}

Similarly, after the third step, the total state is
\begin{eqnarray} \label{3rd01}
|\psi_{3}\rangle &=& {1 \over 2\sqrt{2}} \big( | 0 \rangle_{\rm
int} \otimes \nonumber \\ && \left[ {\rm e}^{\rm i {\pi \over 4}}
| 3 \rangle_{x} + ({\rm e}^{\rm i {\pi \over 4}}+ {\rm e}^{-\rm i
{\pi \over 4}}) | 1 \rangle_{x} - {\rm e}^{- \rm i {\pi
\over 4}} | -1 \rangle_{x} \right] \nonumber \\
&& ~~~+ |1 \rangle_{\rm int} \otimes \nonumber \\ && \left[ {\rm
e}^{\rm i {\pi \over 4}} | 1 \rangle_{x} - ({\rm e}^{\rm i {\pi
\over 4}}- {\rm e}^{-\rm i {\pi \over 4}})| -1 \rangle_{x} + {\rm
e}^{- \rm i
{\pi \over 4}}| -3 \rangle_{x}\right] \big). \nonumber \\
\end{eqnarray}
This demonstrates that the interference of each atomic state ($|
0\rangle_{\rm int}$ and $| 1 \rangle_{\rm int}$) occurs
respectively at position $x=1$ and $-1$ during quantum walks.
Thus, the quantum walk is achieved by alternating superlattices,
and produce the desired probability distribution of an atom over
the sites of the 1D optical lattice.

\section{Physical Implementation}
\label{Physical}

Let us now consider how to implement each of the elements
necessary for realizing quantum walks in an 1D optical lattice.
The required trapping of a single atom in the optical lattice can
be achieved in a variety of ways. Subsequently, alternating
superlattices are required to achieve state-dependent quantum
walks. The realization of 1D and 2D superlattices have been
recently demonstrated~\cite{Sebby-Strabley,Phillips}. We analyze
the amplitudes of the lasers necessary to generate the desired
tunneling ratios. Furthermore, the condition is derived for
performing the optical lattice modulations adiabatically to avoid
heating the trapped atom. We investigate how robust the result of
quantum walks is when imperfect tunneling is considered. Finally,
we describe a measurement scheme that can distinguish the behavior
of quantum walks from their corresponding classical counterpart.

\begin{figure}[t]
\centering
\begin{minipage}{\columnwidth}
\begin{center}
\resizebox{\columnwidth}{!}{\rotatebox{0}{\includegraphics{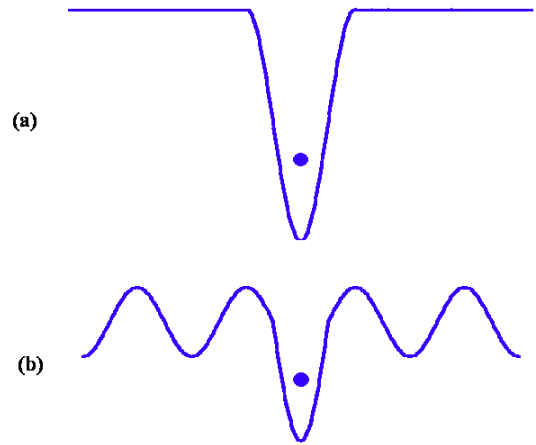}
}}
\end{center}
\vspace{0cm}  \caption{ \label{P_fig01} (a) A single atom is
trapped in a position and (b) the trapping potential is reduced
and a 1D optical lattice is employed.}
\end{minipage}
\end{figure}

\subsection{Single atom trapping in 1D optical lattice}
\label{Howto}

In order to trap a neutral atom, a simple method is to produce a
deep harmonic potential. A well-focused laser beam builds a
micro-trap (an optical tweezer trap) which can grab a single atom
in an optical potential \cite{Pinkse,Darquie}. The schemes of
trapping pointer atoms are also available to generate a single
atom at a certain site \cite{Solano,Kay06}. Alternatively, one can
use the quantum Zeno effect by continuously fluorescent
measurements to obtain a pointer atom \cite{Kay}. When an optical
lattice is applied across the trapping region, the atom can remain
stable in a minimum of optical lattice (see Figure~\ref{P_fig01}).
Thus, to trap a single atom at a certain site within a 1D optical
lattice seems experimentally feasible.

\subsection{Laser amplitude regime}
\label{Parameters} As we have seen, in order to perform step
operation in the quantum walk one needs to control the laser
amplitude corresponding to each atomic state. This will result to
a tunneling coupling $J_\sigma$ that depends on the internal
state, $\sigma$, of the atom. In particular, the tunneling
couplings depend on the amplitude of the laser radiation, $V$, in
the following way~\cite{Scheel}
\begin{eqnarray} \label{HopE}
J ~(V) ={E_R \over 2} \exp (- {\pi^2 \over 4} \sqrt{ V \over E_R})
\left[\sqrt{V \over E_R} + \left( {V \over E_R}
\right)^{3/2}\right],~~
\end{eqnarray}
where $E_R$ is the recoil energy.

\begin{figure}[t]
\centering
\begin{minipage}{\columnwidth}
\begin{center}
\resizebox{\columnwidth}{!}{\rotatebox{0}{\includegraphics{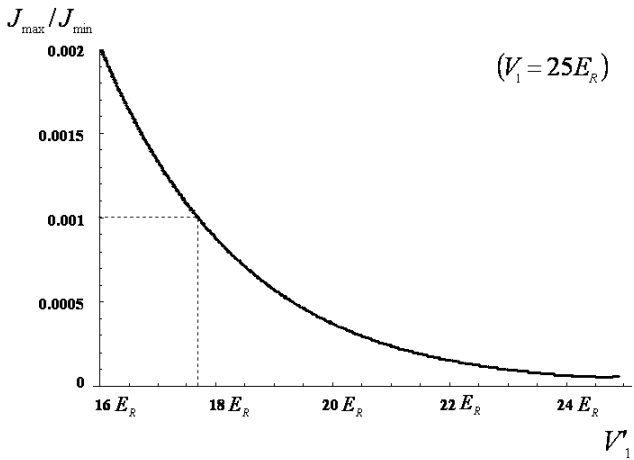}
}}
\end{center}
\vspace{0cm} \caption{ \label{Ratio_J} This demonstrates that more
perfect tunneling is produced for fixed laser coupling $V_1$ when
the second laser coupling $V'_1$ becomes larger.}
\end{minipage}
\end{figure}

Hence, by selectively varying the amplitudes of the laser
radiation at the in-between site regions it is possible to
activate the tunneling interaction between the desired lattice
sites. This is achieved by superlattices that give the spatial
amplitude variation $V_{\rm max} = V_{i}+V'_{i}$ and $V_{\rm min}
=V_{i}-V'_{i}$ ($i=0,1$). If the tunneling of the maximal
potential $V_{\rm max}$ is much smaller than that of the minimal
potential $V_{\rm min}$, the result of our quantum walk achieves
the required outcome. Taking into account the two tunneling
couplings $J_{\rm max} = J ~(V_{\rm max})$ and $J_{\rm min} = J
~(V_{\rm min})$, we demand that the ratio $J_{\rm max} / J_{\rm
min}$ is close to zero. For instance, the amplitude of the first
optical lattice $V_0$ in Eq. (\ref{OL01}) can be taken to be
approximately $25\,E_R$, i.e. the system is taken well into the
Mott-insulating phase \cite{Mandel}. When the second optical
lattice with amplitude $V'_0$ is applied, we consider $J_{\rm max}
/ J_{\rm min}$ as a function of $V'_0$. As shown in Figure
\ref{Ratio_J}, when $V'_0 \approx 17.5\,E_R$, the ratio of two
couplings can be maintained at a sufficiently small value (e.g.
$J_{\rm max} / J_{\rm min} \approx 0.001$) to suppress undesired
tunneling.

In addition, the superlattices must be switched on adiabatically.
Typically, an atom in Mott-insulating optical lattices is trapped
in a ground state. If we change optical potentials rapidly the
atom can be kicked to an excited state or out of the optical
potentials. Thus, adiabatic modulation from optical lattices to
superlattices is required to avoid heating the trapped atom. For
the experimentally achievable trapping frequency of $\omega_{\rm
T}=30$kHz \cite{Mandel, Jaksch98} a suitable time $\delta T\approx
33 \mu s$ can be employed for the adiabatic evolution.

\subsection{Study of errors}
\label{errors}

\begin{figure}[b]
\centering
\begin{minipage}{\columnwidth}
\begin{center}
\resizebox{\columnwidth}{!}{\rotatebox{0}{\includegraphics{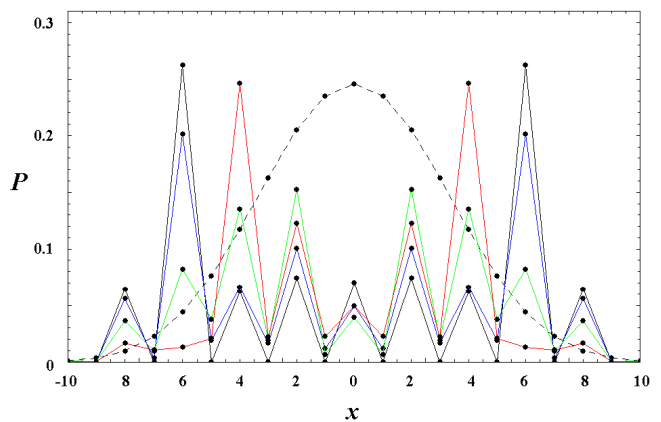}}}
\end{center}
\caption{ \label{4th} Probability distributions for $n=10$ without
errors $\delta t_0 = 0 $ (black) with errors $\delta t_0 = 0.2 /
J_{\rm min}$ (blue), $0.4 / J_{\rm min}$ (green), and $0.6 /
J_{\rm min}$ (red). The dashed line shows the probability
distribution  for $n=10$ in a classical walk.}
\end{minipage}
\end{figure}

Here, sources of experimental errors are taken into account. In
Ref. \cite{Dur}, the decoherence of internal states is mainly
considered because this reduces the quantum walk to its classical
counterpart. The instability of laser beams (e.g. uncontrollable
phase shifts) can influence the mobility of trapped atoms and
cause imperfect manipulations during quantum walks. Nevertheless,
we can restrict ourselves to less than two dozen steps in quantum
walks, where the coherence of the internal states can be
maintained.

To model these errors we consider an incomplete tunneling produced
by the imperfect modulation of trapping potentials. In our setup,
a major error can be produced by the fluctuation of the laser
pulse time $\Delta t=\pi/J(V_\text{min})$. Taking into account
this undesired effect, the unitary operation in Eq. (\ref{U}) is
no longer $\sigma_x$ during the tunneling procedure. When the
tunneling by lowering optical potentials is not perfectly timed,
then the atomic state becomes a superposition state in position
during the tunneling. In this case, quantum walks cannot be
described by the step operator $S$ in Eq. (\ref{S}). Even though
this defect does not cause the decoherence of internal states, it
generates a different kind of quantum walk through odd and even
steps. Shapira {\em et al.} \cite{Shapira} investigated the case
of similar unitary noises in the Hadamard operations during 1D
quantum walks. They showed that the procedure with unitary noise
evolves from quantum to classical walk distributions depending on
the number of steps. Here we consider unitary errors in the step
operations. A more controllable setup can vary the period of laser
pulses in the Hadamard or step operation, thus generating
different step operators and quantum coin tossing, which may be
used for aperiodic quantum walks \cite{Ribeiro}.

As shown in Figure \ref{4th}, the error case demonstrates a
leakage of probability distribution compared with perfect
tunneling and also shows different aspects from the classical
walks. When the period of the laser beam (with error $\delta t_0$)
is described by $\Delta t = \pi / J_{\rm min} + \delta t_0$, the
larger error value of the laser pulse time $\delta t$ generates a
narrower dispersion of the atomic states (see color lines in
Figure \ref{4th}). Eventually, if $\delta t = \pi / J_{\rm min}$,
the characteristic interference behavior of quantum walks does not
appear due to the lack of a step operation in the procedure.

\subsection{Measurement procedure}
\label{Measurements}
\begin{figure}[t]
\centering
\begin{minipage}{\columnwidth}
\begin{center}
\resizebox{\columnwidth}{!}{\rotatebox{0}{\includegraphics{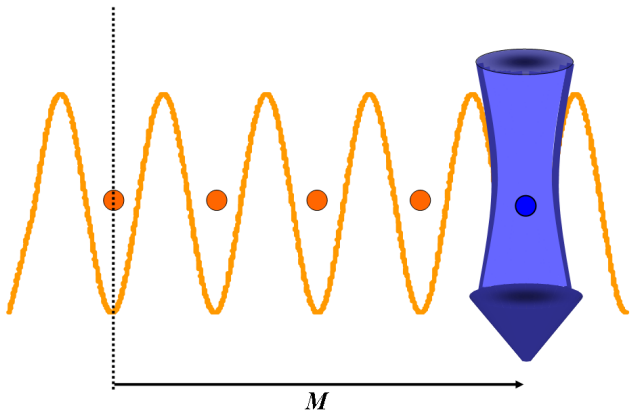}}}
\end{center}
\vspace{-1cm} \caption{ \label{5th} Fluorescence measurements by a
short wavelength laser beam at distance $M$ to obtain a
probability distribution.}
\end{minipage}
\end{figure}

In order to make sure that quantum walks have been performed, we
need to evaluate the probability distribution at a certain lattice
position. When we use relatively long-wavelength optical lattices,
a single atom can be measured at a certain site at distance $M$
from the initial site by a well-focused laser beam using
fluorescent measurements \cite{Roos} (see Figure \ref{5th}).

If we want to measure the atomic state in the $N$-th site away
from the initial site, we shift the measurement laser to the
distance $M=N \lambda$ ($\lambda$ is the wavelength of the
trapping laser). Then, the laser pulse is applied continuously for
a certain period. Here we can assume that the width of the focused
laser beam sufficiently small enough not to disturb the other
states in neighboring sites. By observing fluorescence histograms,
the atomic state in the site can be measured \cite{Roos}.

In this way one can distinguish between a quantum walk evolution
and a classical one by detecting at which step, $n$, of the
control procedure, population has been built at a site $N$. If one
measures the atomic state over all sites in each step, the
distribution of probabilities in a certain step shows the behavior
characteristic of classical walks. Then, the probability at a
specific site increases with respect to more steps due to the
dispersion of the atomic state. However, if we only measure the
atomic state at the end of the quantum walks, the  occupation
probability at the specific site fluctuates over the different
number of total steps.

\begin{figure}[h]
\centering
\begin{minipage}{\columnwidth}
\begin{center}
\resizebox{\columnwidth}{!}{\rotatebox{0}{\includegraphics{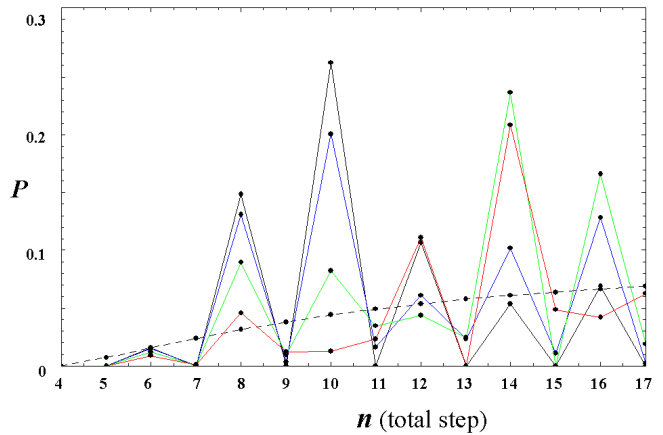}}}
\end{center}
\vspace{0cm} \caption{ \label{6th} The plot shows that the
probabilities of quantum walks (lined) at the sixth site ($N=6$)
fluctuates between the fourth and seventeenth steps while that of
classical walks (dashed) increases continuously. (errors with the
same colors as shown in Figure \ref{4th}).}
\end{minipage}
\end{figure}

In Figure \ref{6th}, we plot the probability at the same site in quantum
walks with respect to the number of steps. The probability at the sixth site
appears after the fifth step in both classical and quantum cases. While the
value of probabilities gradually grow up in the classical case over lager
$n$, the probabilities rapidly fluctuate in the quantum walk cases  due to
quantum coherence. For example, the probability at the tenth step ($n=10$)
in perfect quantum walks reaches the highest peak, approximately 0.26, and
drops to zero while it only increases monotonically from about 0.04 in the
classical distribution. Therefore, we see that a reliable characterization
of the evolution can be deduced within several steps of quantum walks.

\section{Conclusion}

In this article we have proposed a physical realization of a
quantum walk with one atom manipulated within time varying optical
lattices. A detailed study of the required elements is given and a
study of the relevant errors is presented. The interaction
coupling of the evolution is given by direct tunneling of an atom
from one site to its neighbor which can be made significantly
large. Imperfect tunneling generates unitary errors in quantum
walks but it still shows quantum behaviors in probability
distribution. Finally, we describe a feasible measurement approach
which can distinguish the two different probability distributions
of classical and quantum walks at a certain site. This gives the
possibility to perform several algorithmic steps within the
decoherence times of the system, typically taken to be of the
order of a second \cite{Greiner}.

\section{Acknowledgment}
J. J. is supported by an IT Scholarship from the Ministry of
Information and Communication, Republic of Korea and the Overseas
Research Student Award Program. This work was also supported in
part by the European Union Networks SCALA and CONQUEST and by the
UK Engineering and Physical Sciences Research Council
Interdisciplinary Research Collaboration on Quantum Information
Processing.


\begin{thebibliography}{9}

\bibitem{Aharonov}
Y. Aharonov, L. Davidovich, and N. Zagury, Phys. Rev. A {\bf 48},
1687 (1993).

\bibitem{Knight1} P. L. Knight, E. Roldan, and J. E. Sipe,
Phys. Rev. A {\bf 68}, 020301(R) (2003); M. Hillery, J. Bergou,
and E. Feldman, Phys. Rev. A {\bf 68}, 032314 (2003); H. Jeong, M.
Paternostro, and M. S. Kim, Phys. Rev. A {\bf 69}, 012310 (2004);
K. Eckert, J. Mompart, G. Birkl, and M. Lewenstein, Phys. Rev. A
{\bf 72}, 012327 (2005).

\bibitem{Dur}
W. D\"ur, R. Raussendorf, V. M. Kendon, and H.-J. Briegel, \pra
{\bf 66}, 052319 (2002).

\bibitem{Ma}
Z.-Y. Ma and K. Burnett, M. B. d¡¯Arcy, and S. A. Gardiner, Phys.
Rev. A {\bf 73}, 013401 (2006); B. C. Sanders, S. D. Bartlett, B.
Tregenna, and P. L. Knight, Phys. Rev. A 67, 042305 (2003); B. C.
Travaglione and G. J. Milburn, Phys. Rev. A {\bf 65}, 032310
(2002).

\bibitem{Jaksch04}
D. Jaksch,  Contemp. Phys. {\bf 45}, 367 (2004).

\bibitem{Kay}
A. Kay and J. K. Pachos, New J. Phys. 6 126 (2004);
quant-ph/0406073.

\bibitem{rev}
T. P. Spiller, W. J. Munro, S. D. Barrett, and P. Kok, Contemp.
Phys. {\bf 46}, 407 (2005) and references therein.

\bibitem{Phillips}
S. Peil, J. V. Porto, B. L. Tolra, J. M. Obrecht, B. E. King, M.
Subbotin, S. L. Rolston, and W. D. Phillips, Phys. Rev. A {\bf
67}, 051603(R) (2003).

\bibitem{Sebby-Strabley}
J. Sebby-Strabley, M. Anderlini, P. S. Jessen, and J. V. Porto,
Phys. Rev. A {\bf 73}, 033605 (2006).

\bibitem{Pinkse}
P. W. H. Pinkse, T. Fischer, P. Maunz, and G. Rempe, Nature {\bf
404}, 365 (2000).

\bibitem{Darquie} B. Darquie, M. P.
A. Jones, J. Dingjan, J. Beugnon, S. Bergamini, Y. Sortais, G.
Messin, A. Browaeys, and P. Grangier, Science {\bf 309}, 454
(2005).

\bibitem{Solano}
K. G. H. Vollbrecht, E. Solano, and J. I. Cirac, Phys. Rev. Lett.
{\bf 93}, 220502 (2004).

\bibitem{Kay06}
A. Kay, J. Pachos, and C. S. Adams, Phys. Rev. A {\bf 73}, 022310
(2006).

\bibitem{Scheel}
S. Scheel, J. K. Pachos, E. A. Hinds, and P. L. Knight, Lect.
Notes Phys. {\bf 689}, 47 (2006); quant-ph/0403152.

\bibitem{Mandel}
O. Mandel, M. Greiner, A. Widera, T. Rom, T. W. H\"ansch, and I.
Bloch, Nature {\bf 425}, 937 (2003).

\bibitem{Jaksch98} D. Jaksch, C. Bruder, J. I. Cirac, C. W. Gardiner, and P. Zoller \prl {\bf 81}, 3108
(1998).

\bibitem{Shapira}
D. Shapira, O. Biham, A. J. Bracken, and M. Hackett, Phys. Rev. A
{\bf 68}, 062315 (2003).

\bibitem{Ribeiro}
P. Ribeiro, P. Milman, and R. Mosseri, \prl {\bf 93}, 190503
(2004).

\bibitem{Roos}
C. F. Roos, M. Riebe, H. H\"affner, W. H\"ansel, J. Benhelm, G. P.
T. Lancaster, C. Becher, F. Schmidt-Kaler, and R. Blatt, Science
{\bf 304}, 1478 (2004).

\bibitem{Greiner}
M. Greiner, O. Mandel, T. Esslinger, T. W. H\"ansch, I. Bloch,
Nature {\bf 415}, 39 (2002).


\end{thebibliography}
\end{document}